\documentclass[%
 notitlepage,
 reprint,
 superscriptaddress,
 amsmath,amssymb,
 aps,
]{revtex4-1}

\usepackage{lipsum}
\usepackage{graphicx}
\usepackage{dcolumn}
\usepackage{bm}

\usepackage{epstopdf}

\newcommand{\bra}[1]{\langle #1 \vert}
\newcommand{\ket}[1]{\vert #1 \rangle}

\usepackage{hyperref}
\usepackage{xcolor}

\begin{document}

\title{Localization properties of a two-channel 3D Anderson model}

\author{Andre M. C. Souza}
\affiliation{%
 Departamento de F\'{i}sica, Universidade Federal de Sergipe, 49100-000 S\~{a}o Crist\'{o}v\~{a}o, SE, Brazil
}%

\author{Guilherme M. A. Almeida}
\email{gmaalmeida@fis.ufal.br}
\affiliation{%
 Instituto de F\'{i}sica, Universidade Federal de Alagoas, 57072-900 Macei\'{o}, AL, Brazil
}%

\author{Eduardo R. Mucciolo}
\affiliation{%
 Department of Physics, University of Central Florida, Orlando, FL 32816, USA
}%

\date{\today}

\begin{abstract}
We study two coupled 3D lattices, one of them featuring uncorrelated on-site disorder and the other one being fully ordered, 
and analyze how the interlattice hopping affects the localization-delocalization transition of the former and how
the latter responds to it. We find that moderate
hopping pushes down the critical disorder strength for the disordered channel
throughout the entire spectrum compared to the usual phase diagram for the 3D Anderson model. 
In that case, the ordered channel begins to feature an effective disorder also leading to 
the emergence of mobility edges but with higher associated critical disorder values. 
Both channels become pretty much alike as their hopping strength is further increased, as expected.
We also consider the case of two disordered components and show that in the presence of certain correlations among the parameters of both lattices, one obtains a disorder-free channel decoupled from the rest of the system.   
\end{abstract}

\maketitle

\section{\label{sec1}Introduction}

Put forward many decades ago and named after its discoverer, Anderson localization is 
one of the most groundbreaking outcomes in condensed matter physics \cite{anderson58}, having been
covered in a wide context in recent years \cite{evers08}. 
In a nutshell, it implies that the wavefunction of
noninteracting quantum particles 
becomes trapped around a finite
region of 1D and 2D lattices given
any amount of randomness in the on-site energy distribution \cite{abrahams79}, what
dramatically affects the transport properties of the system. 
In 3D (and higher-dimensional) lattices, there is a
localization-delocalization transition
for critical values of the disorder strength,
with well defined mobility edges \cite{abrahams79}.
Many experiments performed on ultracold atoms have
characterized such transition \cite{kondov11, jendrzejewski12, semeghini15}.

Things get even more involved when disorder happens to feature embedded positional
correlations. Early works showed that short-range correlations 
are capable of inducing extended states in 1D \cite{flores89, dunlap90},
while long-range-correlated disorder was found to support a continuous band
of extended states in the middle of the band, featuring
an Anderson-type transition with sharp mobility edges \cite{demoura98, izrailev99}.
An extended phase was also reported in a 2D disordered model featuring correlated impurities \cite{hilke03}. Coexistence between localized and extended states has also been
addressed for a tight-binding model involving electron-mass position dependence \cite{souza19}.

Another class of low-dimensional disordered models that 
has been enjoying a great deal of attention 
is that of ladderlike (laterally-coupled) disordered chains 
\cite{heinrichs02, heinrichs03, romer04, sedrakyan04, roemer05, diaz07, bagci07, sil08prl, sil08prb, demoura10, zhang10, guo11, rodriguez12, xie12,
nguyen12, zhao12, guo13, weinmann14, bordia16, mastropietro17, zhao17, almeida18DFS}, 
traditionally used
for studying electronic transport in 
double-stranded DNA molecules (see, e.g., \cite{roemer05, diaz07, bagci07}).
In \cite{sil08prl} it was reported that a ladder made of two coupled Aubry-Andr\'{e} chains
displays a metal-insulator at multiple Fermi-energy levels. 
Shortly after, it was shown in Ref. \cite{sil08prb} (see also \cite{demoura10}) that two-leg random ladders may
exhibit a band of Bloch-type extended states provided the on-site potentials and 
interchain coupling strengths obey a set of correlations.
The emergence of such disorder-free subspaces was generalized to many-leg ladders in \cite{rodriguez12} and \cite{guo13}, the latter for 
a random binary layer model, and may find applications in quantum information processing as well \cite{almeida18DFS}. 
%

Coupled lattices also emerge, in an effective way, 
when spin-orbit coupling is taken into account in the Anderson model \cite{asada02}. 
In such, electron transport is
affected by its intrinsic angular momentum as
spin-up and spin-down channels are now coupled, what adds another dimension to the problem.
Interest in this class of models (featuring broken SU(2) symmetry \cite{evers08}) burst
with the findings that inclusion of spin-orbit coupling allows
for an Anderson transition in 2D \cite{hikami80}. 
This have been investigated numerically 
on various settings \cite{asada02, evangelou87,ando89, merkt98, orso17}, including noninteracting particles with higher spins \cite{su18}.
An accurate estimate of the critical exponent $\nu$ associated to the localization-length divergence can be found in \cite{asada02} for the
symplectic university class.
Physical implementations in optical lattices have also been discussed \cite{orso17, zhou13}
as significant progress has been made in tuning synthetic spin-orbit coupling for cold atoms 
\cite{huang16}.

Some interest has also been directed toward ladder models
featuring channels with different degrees and/or types of disorder \cite{zhang10, guo11, zhao12,guo13}.
For instance, Zhang \textit{et al.} \cite{zhang10} addressed the case of an
ordered chain coupled to a disordered one and reported that every eigenstate of the
system becomes outright localized given the disorder is uncorrelated, 
even though particle transport is enhanced (suppressed) in the disordered (ordered) component.  
They also investigated the case of long-range correlated disorder and described 
a quantum phase transition taking place at a critical interchain
coupling strength.      
Two-channel models also find support in the context of polaritons, e.g. mixed particles of light and matter, 
where each component may come with different degrees of disorder due to their very
nature \cite{xie12}. 

Thus a bilayered graph, in general, may be spanned intrinsically (such as in the presence of spin-orbit coupling) or not. 
As far as Anderson localization is concerned though, what matter the most are  
the system dimensionality and its underlying symmetries given the phenomena 
is primarily driven by interference. 
In this work, we aim to address the interplay 
between an Anderson Hamiltonian and an integrable one. 
More specifically, we investigate the localization properties arising
from coupling two 3D simple-cubic lattices (see Fig. \ref{fig1}), 
one being an ordered channel and the other one featuring on-site uncorrelated disorder.
Note that our system can simply be seen as 
a 3D Anderson model featuring a two-level system per site.
The main goal here is to find out, from each channel's point of view, 
how the coupling affects the localization-delocalization phase diagram of the disordered channel as well as
how this transition takes place in the (hitherto) ordered channel.
We do that by evaluating the participation ratio properly defined for each channel. 
We find out that moderate interchannel hopping strength, while 
decreasing the critical disorder strengths associated to 
the disordered channel only, does not necessarily lead
the full (bilayered) lattice to a localized phase as 
one is able to find delocalized states in the middle of the spectrum band which is
mostly dominated by eigenstates overlapping with the ordered channel.
Furthermore, we deal with two disordered channels 
with correlated parameters in order to span a disorder-free channel 
thereby extending the framework made for the 1D ladder framework \cite{sil08prb}
to 3D.

This work is organized as follows. In Sec. \ref{sec2} we introduce the Hamiltonian model for the 3D bilayered lattice. Then in Sec. \ref{sec3} we 
evaluate the localization properties of the system via exact numerical diagonalization
and analyze its phase diagrams. Following that we work out analytically 
the requirements for generating an uncoupled ordered channel out of two coupled 3D disordered channels in
Sec. \ref{sec4}. Conclusions are drawn in Sec. \ref{sec5}

\section{\label{sec2}Model}

\begin{figure}[t!] 
\includegraphics[width=0.45\textwidth]{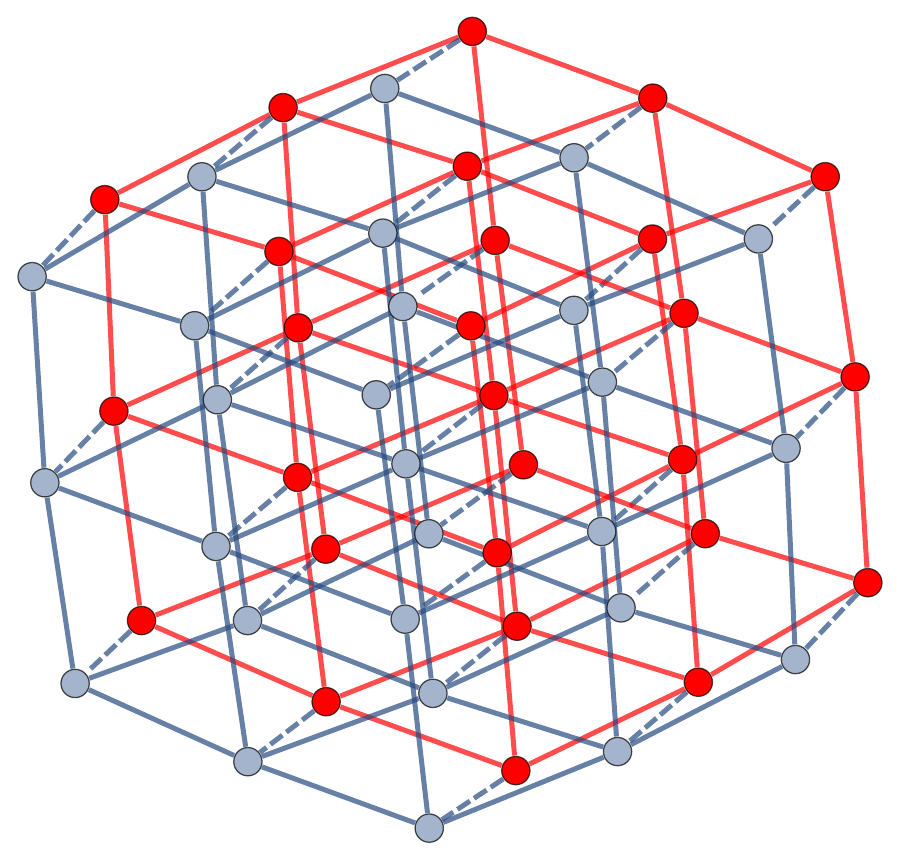}
\caption{\label{fig1} Two-channel 3D Anderson lattice (with $L=3$). An ordered system 
is coupled to a fully disordered one via $U$ (dashed edges). The latter features
on-site potentials $\epsilon_{i}$ falling randomly within $[-W/2,W/2]$ [say, red (dark) vertices].
Here, we set the intra-lattice couplings (solid edges) to $t \equiv 1$. Note that the full Hilbert space
is $2N$-dimensional, with $N = L^{3}$.
}
\end{figure}

The Hamiltonian describing a single particle (e.g. an electron) hopping through a 3D bilayered lattice, with $N =L^{3}$ sites each,  
reads $\hat{H} = \hat{H}_{1}+\hat{H}_{2}+\hat{H}_{I}$, where 
\begin{equation} \label{H_layer}
\hat{H}_{\ell} = \sum_{i}\epsilon_{\ell,i} \hat{c}_{\ell, i}^{\dagger}\hat{c}_{\ell, i} + t\sum_{\langle i,j \rangle}\left( \hat{c}_{\ell, i}^{\dagger}\hat{c}_{\ell, j} + \mathrm{H.c.} \right)
\end{equation}
is the local tight-binding Hamiltonian for each layer ($\ell = 1,2$), with $t$ and $\epsilon_{\ell, i}$ being, respectively, 
the intra-hopping strength and the on-site potential, and $\hat{c}_{\ell, i}$ ($\hat{c}_{\ell, i}^{\dagger}$) denoting the fermionic
annihilation (creation) operator at site $i$ of the $\ell$-th layer. The sum in the second term of Eq. (\ref{H_layer})
runs over nearest-neighbors sites of a simple cubic lattice.
The coupling between both channels is accounted by 
\begin{equation} \label{H_int}
\hat{H}_{I} = \sum_{i}U_{i}\left( \hat{c}_{1, i}^{\dagger}\hat{c}_{2, i} + \mathrm{H.c.} \right),
\end{equation}
where $U_{i}$ is the inter-layer hopping strength.

At this point, we are to make a few assumptions towards the parameters of the system. First, note that $t$ is constant for both
layers, and we set it as our energy unit ($t\equiv 1$). For now, we also set $U_{i} = U$ uniform and assume that
one of the layers features no disorder at all, with $\epsilon_{1,i} = 0$ whereas $\epsilon_{2,i}$ is taken
out of a box distribution within $[-W/2,W/2]$, $W$ being the disorder strength. 
This configuration is depicted in Fig. \ref{fig1}.

For $U=0$, both layers are decoupled. The ordered one features extended wave functions over the whole the energy spectrum. 
On the other hand, layer 2 itself is a 3D Anderson model for which
a transition between extended and localized states takes place for a given critical value of disorder strength $W_c$ that depends on the energy level and lattice topology. 
For instance, $W_{c}=0$ for 1D and 2D at any energy level, meaning
that every eigenstate of the system is exponentially localized even in the presence of the tiniest amount of disorder \cite{abrahams79}.
For higher dimensions, say, in a simple cubic lattice, the transition is found for $W_c / t \approx 16.53$ at the middle of the band ($E=0$) \cite{slevin14}. 
%
The mobility edge, 
that is the critical energy above which the particles are free to move 
has been estimated using ultracold atoms in optical lattices \cite{semeghini15}.

\section{\label{sec3}Localization properties}

Hereafter we are interested in the case $U\neq 0$.
Were both layers ordered (say, setting $\epsilon_{2,i} = 0$ as well), things would be straightforward to deal with.
In this particular scenario, the Hamiltonian can be handled out analytically 
in Fourier space, that is
$\hat{H} |\psi_{\vec{k},\sigma} \rangle  = E_{\vec{k},\sigma} |\psi_{\vec{k},\sigma} \rangle$
where the eigenvalues are
\begin{equation} \label{blochen}
E_{\vec{k},\sigma}= \sigma U  + \frac{1}{N} \sum_{\langle i,j \rangle} e^{i\vec{k}\cdot (\vec{R}_{i}-\vec{R}_{j})}  ,
\end{equation} 
$\sigma = \pm 1$, $\vec{R}_{i}$ is the position vector of the $i$-th vertice in the lattice, and $\vec{k}$ is the reciprocal lattice vector satisfying $\vec{k} \cdot \vec{R}_{i} = 2\pi n$, where $n$ is an integer. The eigenvectors are
\begin{equation} \label{blocheigen}
|\psi_{\vec{k},\sigma} \rangle  = \frac{1}{2N} \sum_{i=1}^{N} e^{i\vec{k} \cdot \vec{R}_{i}} \left( \hat{c}_{1,i}^{\dag} + \sigma \hat{c}_{2,i}^{\dag} \right)  |\emptyset\rangle ,
\end{equation} 
with $|\emptyset \rangle$ being the vacuum state
and $\ket{i}_{\ell} = \hat{c}_{\ell,i}^{\dag} |\emptyset\rangle$
denoting a single particle located at the $i$-th site
of the $\ell$-th channel. Given Eq. (\ref{blocheigen}) it is readily seen that for any eigenstate
the probability of finding the particle at a given location
is $1/2N$, due to the extended character of the Bloch wavefunctions.

Basically, when linking up two identical ordered systems like discussed above, one gets two effective \textit{uncoupled} lattices, with local energies $+U$ and $-U$, respectively,  
and featuring the same dispersion profile [see Eq. (\ref{blochen})]. The states that form those effective structures are even symmetric and anti-symmetric combinations 
of $\ket{i}_{1}$ and $\ket{i}_{2}$, and then there are
two bands of propagating modes at our disposal. At the end of this paper, 
we show that a band of extended states can be activated
even when we couple two disordered lattices,
as long as their parameters obey a certain class of correlations.

Our main goal now is to see about how the 
coupling between the ordered ($\epsilon_{1,i} = 0$) and disordered ($\epsilon_{2,i} \in [-W/2,W/2]$) lattices affects the localization-delocalization transition of the full system.
At a first glance, we are led to think 
that the channels may push each other out, meaning that
transport in the ordered (disordered) lattice is suppressed (enhanced). At least, that is 
the outcome for a 1D ladder chain in the case of uncorrelated disorder \cite{zhang10}. 
Here, however, things should get more involved as the 3D Anderson model features 
mobility edges.
To get into further details regarding our bilayered 3D model, we will resort to exact numerical diagonalization of the Hamiltonian in order to obtain the quantities of interest.  

Let $\ket{\psi_{k}} = \sum_{i=1}^{N}\left( a_{k,i}\ket{i}_{1}+b_{k,i}\ket{i}_{2} \right)$ be the eigenstate associated to level $E_{k}$. Thus, the probability to find the particle in lattice 1 (lattice 2), that is the ordered (disordered) channel, for a given energy level, 
is given by $P_{1}(E_{k})=\sum_{i}|a_{k,i}|^{2}$ ($P_{2}(E_{k})=\sum_{i}|b_{k,i}|^{2}$). Observe that $P_{1}(E_{k})+P_{2}(E_{k}) = 1$. 
The degree of localization can be characterized through
the participation ratio, here defined as
\begin{align} \label{iprc}
R_{1}(E_{k}) &= \frac{ P_{1}(E_{k}) }{ \sum_{i=1}^{N} |a_{k,i}|^{4} },\\
 \label{iprf}
R_{2}(E_{k}) &= \frac{ P_{2}(E_{k}) }{ \sum_{i=1}^{N} |b_{k,i}|^{4} }.
\end{align}
A localized eigenstate is characterized by $R_{\ell}(E_{k})/N \rightarrow 0$ in the limit $N \rightarrow \infty$ and 
the ratio converges to a finite value if the the state happens to be extended.
In what follows, the energy band is divided into twenty intervals, $(\mathrm{max}\lbrace E_{k} \rbrace-\mathrm{min}\lbrace E_{k} \rbrace)/20$, so that the quantities $P_{\ell}(E)$ and $R_{\ell}(E)$ are 
averages taken over the states allocated 
within a small window around $E$.
We further average them out over $m$ independent samples for each chosen values of $L$, $U$, and $W$. 


\begin{figure}
\centering
	\includegraphics[width=0.5\textwidth]{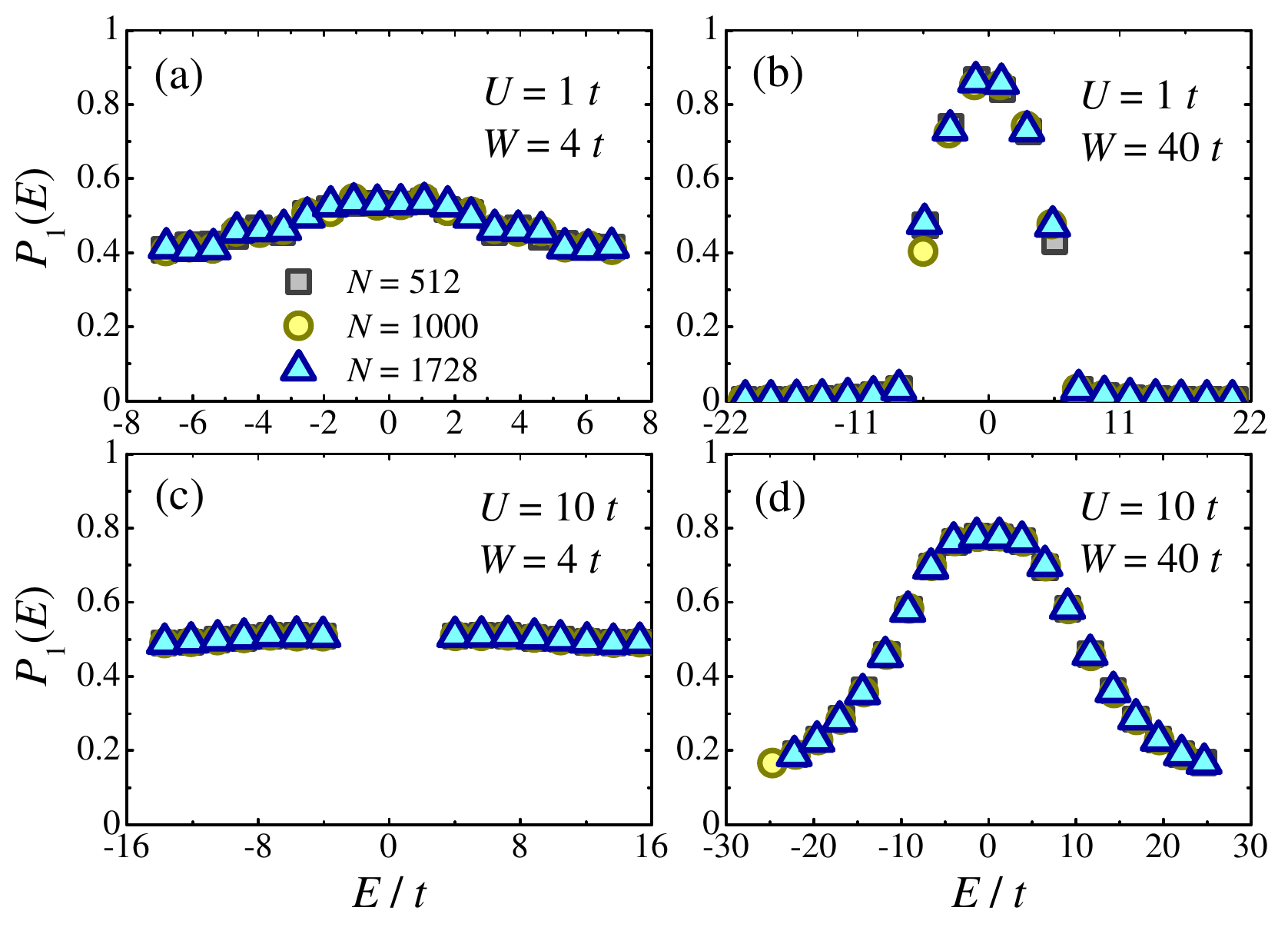}
 \caption{\label{fig2} Probability of finding the particle at lattice 1 (ordered), $P_{1}(E)$, 
averaged over a set of eigenstates surrounding $E$ for (a,b) $U=1t$ and (c,d) $U=10t$, with disorder strengths
$W=4t$ and $W=40t$ in each case. 
Results are obtained for lattice sizes $N=512$ ($L=8$), $N=1000$ ($L=10$), and $N=1728$ ($L=12$), averaged over $m=40$, $20$, and $12$ independent realizations of disorder, respectively.} 
\end{figure}

We start off discussing the overall occupation probability for one of the lattices, say $P_{1}(E)$, accounting for the ordered channel. 
Results are shown in Fig. \ref{fig2} for $U=1t$ and $U=10t$, considering two disorder strengths and different system sizes $N$.
From Eq. (\ref{blocheigen}), valid for $U\neq0$ and $W=0$, we get the idea that in the weak disorder regime ($W \ll U$)
there should still be likely to find the particle in 
any of the channels with almost equal probability.
Already for $W=4t$, though, and interchannel hopping strength $U=1t$,
one starts noticing that the outskirts of the band become slightly less involved with the ordered channel [see Fig. \ref{fig2}(a)]. This reaches a
serious level in the presence of intense disorder [Fig. \ref{fig2}(b)] to the point 
the very center of the band is almost entirely populated by the ordered lattice whilst
the disordered counterpart dominates for higher energies (in absolute values), as imposed by $W$.   
In other words, the eigenstates tend to be no longer mixed upon increasing $W$, 
except for those lying in between as we depart from the middle of the band. 
In general, those properties above still stand for higher hopping strengths, as displayed in Fig. \ref{fig2} for $U=10t$.
However, note in Fig. \ref{fig2}(c) that there is an energy gap 
as $U$ mixes the channels up a great deal and pushes two subbands apart
[cf. Eq. (\ref{blochen})].  
When $W$ is increased [Fig. \ref{fig2}(d)], the gap is closed as eigenstates 
featuring higher overlaps with the ordered channel once again
move to the center of the band, although the probability distribution over $E$ 
is not so as sharp as we have seen in Fig. \ref{fig2}(b) due to the value of $U$. 
We also mention that all those properties are valid regardless of the system size $2N=2L^{3}$.  
  
The above analysis, while revealing some interesting aspects over the population mixedness of the eigenstates in relation to the coupling between ordered and disordered lattices,    
does not really tell about their localization strength. To do so, 
we must proceed with a finite-size scaling analysis for the participation ratio. 
Considering that $R_{\ell}(E_{k}) \sim N^{\alpha}$, with $\alpha \in [0,1]$, 
the state is said to be completely delocalized when $\alpha = 1$ and localized otherwise. 
(One should bear in mind that it is still possible to 
find several localization profiles for the wavefunctions in that region.)

\begin{figure}
	\centering
	\includegraphics[width=0.40\textwidth]{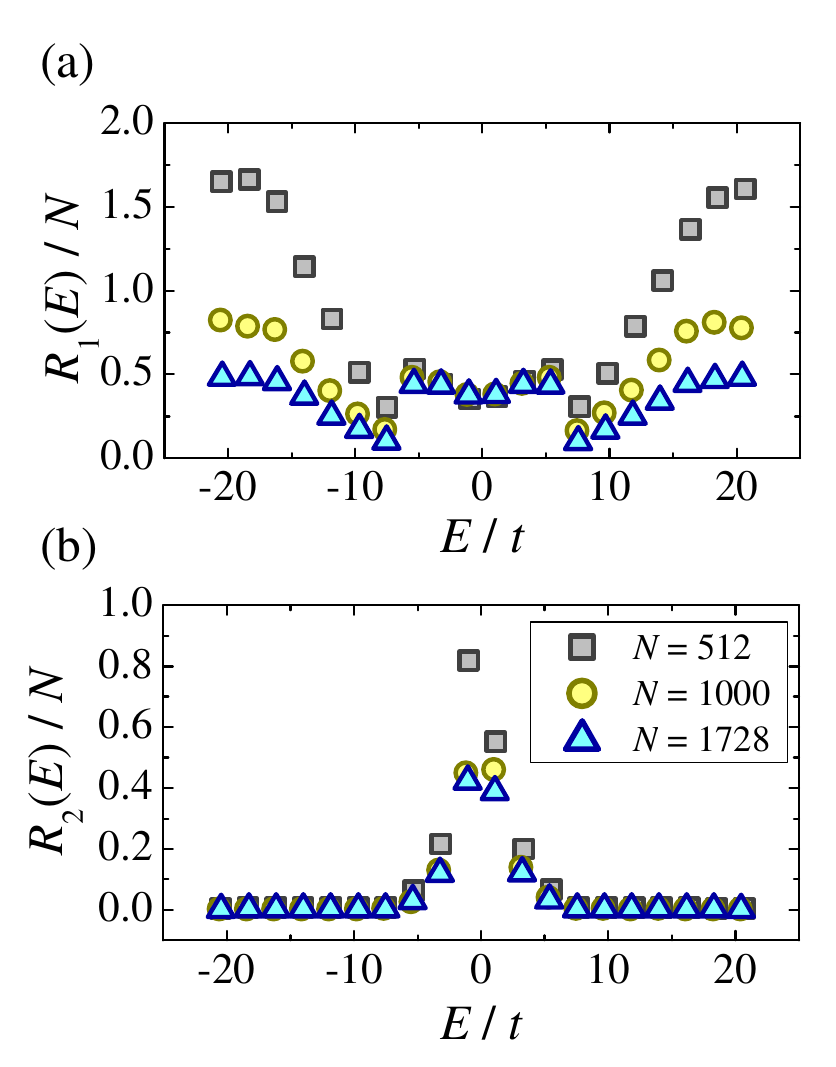}
	\caption{Rescaled participation ratios (a) $R_{1}(E)/N$ (ordered channel) and (b) $R_{2}(E)/N$ (disordered channel) averaged over a set of eigenstates surrounding $E$ for $U=1t$ and $W=40t$.
		Once again, results are shown for lattice sizes $N=512$ ($L=8$), $N=1000$ ($L=10$), and $N=1728$ ($L=12$), averaged over $m=40$, $20$, and $12$ independent realizations of disorder, respectively.
		} \label{fig3}
\end{figure}

\begin{figure}[t!]
\centering
	\includegraphics[width=0.49\textwidth]{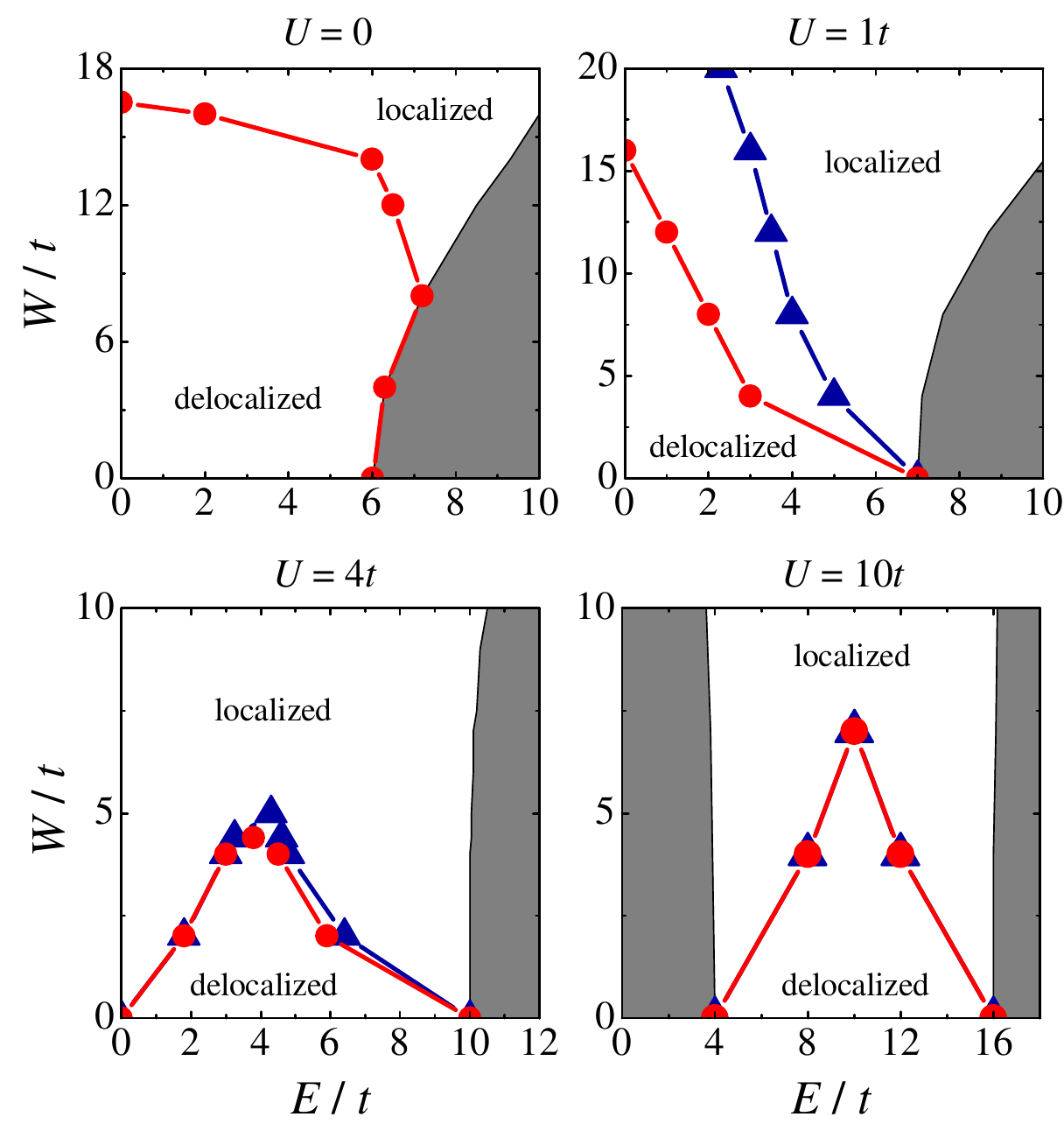}
 \caption{\label{fig4} Localization-delocalization phase diagrams for (ordered) channel 1 (blue triangles) and (disordered) channel 2 (red circles)
for $U/t = 0, 1, 4,$ and $10$ 
evaluated via finite-size scaling analysis of 
$R_{1}(E)$ and $R_{2}(E)$ for $N=512, 1000,$ and $1728$ (averaged over $m=40$, $20$, and $12$ samples, respectively). Gray-shaded areas stand for band gaps. Lines are for guiding the eye.} 
\end{figure}

As an example, in Fig. \ref{fig3} we display the one-lattice participation ratio (divided by $N$), as defined in Eqs.
(\ref{iprc}) and (\ref{iprf}) for $U=1$ and strong disorder ($W=40$), what
eventually enforces localization for both lattices in the entire spectrum.
Therein we have checked that $\alpha< 1$ as expected.
To extract some information over the \emph{degree} of localization though, 
we must look after the value of $R_{\ell}(E_{k})$ itself. 
Although lattice 1 (the ordered channel 1)
is now effectively disordered, its associated coefficients
$\lbrace a_{k,i} \rbrace$ are such that they combine 
to form a set of localized states featuring a much larger
localization length than those associated to 
lattice 2 in the outskirts of the band [where channel 2 dominates (cf. Fig. \ref{fig2}(b))].
In that region, we checked that the corresponding 
$\alpha$ is close to zero 
for both lattices.

We then used the above criteria for the participation ratio to construct
the $W-E$ phase diagrams shown in Fig. \ref{fig4}. 
In the absence of interchannel coupling ($U = 0$), we recover
the standard phase diagram for the 3D Anderson model (which is lattice 2, only) featuring
a critical disorder strength $W_{c}/t$ of about 16.5 at $E=0$, above which
the system is localized.
But then as we set $U/t=1$
thereby connecting lattices 1 and 2, 
the latter becomes more sensitive to disorder
as the reentrant pattern is gone. 
A localization-delocalization transition is also induced in
in channel 1 (hitherto ordered) for higher values of $W_{c}$. 
For intermediate values of $W_{c}$, although the wavefunction components
of lattice 2 suggests a localized phase, lattice 1 still holds
the system in the delocalized phase, especially for energies
around the band center, filled out with eigenstates 
mostly overlapping with the ordered channel.
Both channels happen to feature about the same behavior
when $U/t=4$ due to the band mixing leading to low values of critical disorder strength overall. 
%
For $U/t=10$, the coupling between both lattices is such that one can barely tell 
them apart.
Each subband [notice the gap taking place around the middle of the spectrum; see Fig. \ref{fig2}(c)]  
features a transition around $W_{c}/t = 7$ 
at its center. In this strong-$U$ regime each channel, on its own, is thus able 
to provide valuable information over the whole lattice.

\section{\label{sec4}Disorder-free subspace}

So far we have been dealing with 
the localization properties of a disordered 3D lattice coupled to
a ordered one.  
In this last section, though, we 
consider both lattices to feature on-site disorder and show analytically, following Ref. \cite{sil08prb},
that certain correlations
among parameters $\epsilon_{1,i}$, $\epsilon_{2,i}$, and $U_{i}$ [cf. Eqs. (\ref{H_layer}) and (\ref{H_int})]
can effectively decouple
both channels, 
thereby spanning a \textit{disorder-free} subspace. In the following procedure, 
the intra-lattice hopping strength $t\equiv 1$, still.

Each (two-level-like) cell formed from states $\ket{i}_{1}$ and $\ket{i}_{2}$ features 
the local Hamiltonian
\begin{equation}
H_{i}^{(\mathrm{cell})}=
\begin{pmatrix}
\epsilon_{1,i} & U_{i}\\ 
U_{i} & \epsilon_{2,i}
\end{pmatrix},
\end{equation}
which can be put in diagonal form via
\begin{align}
\ket{+}_{i} &= \sin\theta_i \ket{i}_{1}+\cos\theta_i \ket{i}_{2},\\
\ket{-}_{i} &= \cos\theta_i \ket{i}_{1}-\sin\theta_i \ket{i}_{2},
\end{align}
with correponding eigenvalues 
\begin{equation} \label{E_dressed}
E_{i}^{\pm}=\frac{1}{2}\left(\epsilon_{1,i}+\epsilon_{2,i}\pm \sqrt{4U_{i}^2+\Delta_{i}^{2}} \right),
\end{equation}
where $\Delta_{i} = \epsilon_{2,i}-\epsilon_{1,i}$ is the local energy detuning and 
\begin{equation}\label{theta}
\theta_{i} = \tan^{-1}\left( \frac{2U_{i}}{\sqrt{4U_{i}^2+\Delta_{i}^{2}}+\Delta_{i}} \right).
\end{equation}

Rewriting the system Hamiltonian in terms of 
operators $\alpha_{i}^{(\mu)\dagger} \equiv \ket{\mu}_{i}\bra{\emptyset}$
$(i=1,\ldots,N)$, with $\mu = \pm$, we get 
\begin{align}
H = &\sum_{i,\mu}E_{i}^{\mu}\alpha_{i}^{(\mu)\dagger}\alpha_{i}^{(\mu)}+ \sum_{\langle i,j\rangle, \mu} J_{ij}^{(\mathrm{intra})} \left(\alpha_{i}^{(\mu)\dagger}\alpha_{j}^{(\mu)} + \mathrm{H.c.} \right) \nonumber \\ 
\,\,\,\,\,&+ \sum_{\langle i,j\rangle} J_{ij}^{(\mathrm{inter})} \left( \alpha_{i}^{(+)\dagger}\alpha_{j}^{(-)}  + \mathrm{H.c.} \right),
\end{align}
where
\begin{align}
J_{ij}^{(\mathrm{intra})} &= t \cos (\theta_i - \theta_j),\\ 
J_{ij}^{(\mathrm{inter})} &= t\sin (\theta_i - \theta_j).
\end{align}
The above description keeps the intraconnectivity pattern of each subsystem while 
establishes more interconnections per (effective) site. 
%

If we want one of the 
channels -- that is the positive or the negative branch -- to be free of (diagonal) disorder, the first step is to 
place all of its local energies at the same level, say, zero for simplicity. Then, given $U_{i} = \pm \sqrt{\epsilon_{1,i} \epsilon_{2,i}}$ 
(with $\epsilon_{1,i},\epsilon_{2,i}>0$) one has $E_{i}^{+} =  \epsilon_{1,i}+\epsilon_{2,i}$ and $E_{i}^{-} = 0$ for all $i$ [see Eq. (\ref{E_dressed})].
In this case, we arrive at a similar situation as before, where a disordered lattice (positive branch) is coupled to a ordered one (negative branch). 
Then, we further need to decouple them, by arranging for $J_{ij}^{(\mathrm{inter})} = 0$,
what implies that  $\theta_{i}-\theta_{j} = n\pi$ for any integer $n$. From Eq. (\ref{theta})
we thus see that 
$\tan\theta_{i} = \tan \theta_{j}$ for all $i$ and $j$. 
As, given $U_{i} = \sqrt{\epsilon_{1,i} \epsilon_{2,i}}$ (with no loss of generality), 
$\tan\theta_{i} = 2 \sqrt{\epsilon_{1,i}/\epsilon_{2,i}}$,  
it is then required that $\epsilon_{2,i} / \epsilon_{1,i} = \epsilon$, what makes $U_{i}/ \epsilon_{1,i} = \sqrt{\epsilon}$,
with fixed $\epsilon > 0$ over the entire lattice \cite{sil08prb}. This set of correlations entails 
$J_{ij}^{(\mathrm{intra})} = t$ and, finally, we obtain two independent effective lattices, a clean one and another featuring
on-site disorder with $E_{i}^{+} = \epsilon_{1,i}(1+\epsilon)$, for which results of Anderson localization theory in 3D are known.

\section{\label{sec5}Conclusions}

We explored the localization-delocalization transition in a bilayered (two-channel) Anderson model in three dimensions. First we considered
one of the lattices being ordered, with the other one featuring diagonal uncorrelated disorder, 
and discussed about the role of their coupling upon the mobility edges
of each channel separately, evaluated through the participation ratio. 
In summary, for moderate 
interchannel coupling the ordered channel begins to feature effective on-site fluctuations
leading to relatively weak disorder when compared to the other (already disordered) channel.
Strong coupling leads to mixing between both channels and they thus 
happen to feature almost the same critical disorder strengths along the spectrum. 

Following that, we also considered two coupled disordered 3D lattices and showed how 
to create an effective channel completely free of disorder.
The very coexistence between localized and delocalized states that emerges from
the above and similar frameworks provides with the idea of engineering extended states
in a disordered background \cite{hilke03, sil08prb, rodriguez12, guo13, almeida18DFS}.
While individual chains with correlated disorder already offers a great deal of 
selective transport properties -- that can be used, for instance, in entanglement distribution \cite{almeida17pra, almeida19qip} 
and quantum-state transfer \cite{almeida18ann, junior19} protocols -- 
laterally-coupled channels add more possibilities, given the strength of the interchain hopping as
it is able to mix modes with totally different profiles.

To some extent, this work was also motivated by the fact that recent advances in 
ultracold atoms in optical lattices have reached higher levels of local control \cite{gross17}, not to mention
such platforms had already had success in probing Anderson localization phenomenon in 3D \cite{kondov11, jendrzejewski12, semeghini15}. 
Further extensions of our work should include non-trivial topologies, such 
as complex networks with small-world characteristics for they display 
remarkable localization properties \cite{zhu00, giraud05, cardoso08}. 
It would thus be interesting to see about how concatenating many channels made up of those affects
the dynamical properties of the whole system and evaluate its robustness against disorder and other kinds of noise.

\section*{Acknowledgements}
AMCS thanks financial support of CNPq.

\end{document}